\documentstyle[12pt]{article}
\input psfig
\newcommand{\be}{\begin{equation}}
\newcommand{\ee}{\end{equation}}
\newcommand{\nut}{\nu_\tau}

\newcommand{\MeV}{\rm MeV}

\def\re#1{{[\ref{#1}]}}
\def\ut#1{\mathop{\vtop{\ialign{##\crcr
     $\hfil\displaystyle{#1}\hfil$\crcr\noalign
     {\kern1pt\nointerlineskip}\hbox{$\hfil\sim\hfil$}\crcr
     \noalign{\kern1pt}}}}}
\def\undersim{\ut}
\def\la{\mathrel{\mathpalette\fun <}}

\def\fun#1#2{\lower3.6pt\vbox{\baselineskip0pt\lineskip.9pt
        \ialign{$\mathsurround=0pt#1\hfill##\hfil$\crcr#2\crcr\sim\crcr}}}
\begin{document}
\begin{titlepage}
\null\vspace{-72pt}
\begin{flushright}
{\footnotesize
FERMILAB--Pub--96/061-A\\
UUITP-5/96\\
astro-ph/9603051\\
March 1996 \\
Submitted to {\em Phys.\ Rev.\ D}}
\end{flushright}
\renewcommand{\thefootnote}{\fnsymbol{footnote}}
\vspace{24pt}

\baselineskip=24pt

\begin{center}
{\Large \bf  Cosmological Bounds to the Magnetic\\
                     Moment of Heavy Tau Neutrinos} \\
\baselineskip=14pt
\vspace{24pt}
Dario Grasso\footnote{Electronic address: {\tt grasso@atlas.teorfys.uu.se}}\\
{\em Department of Theoretical Physics, Uppsala University\\
Box 803, S-751 08 Uppsala, Sweden, and \\
Department of Physics, University of Stockholm\\
Vanadisv\"agen 9, S-113 46 Stockholm, Sweden}
\vskip 12pt
Edward W.\ Kolb\footnote{Electronic address: {\tt rocky@rigoletto.fnal.gov}}\\
{\em NASA/Fermilab Astrophysics Center\\
Fermi National Accelerator Laboratory, Batavia, Illinois~~60510, and\\
Department of Astronomy and Astrophysics, Enrico Fermi Institute\\
The University of Chicago, Chicago, Illinois~~ 60637}\\
\end{center}

\baselineskip=24pt

\vspace{12pt}

\begin{quote}
\hspace*{2em} The magnetic moment of tau neutrinos in the MeV mass
range may be large enough to modify the cosmological freeze-out
calculation and determine the tau-neutrino relic density.  In this
paper we revisit such a possibility.  We calculate the evolution and
freeze-out of the tau neutrino number density as a function of its
mass and magnetic moment.  We then determine its relic density, then
calculate its effect upon primordial nucleosynthesis including
previously neglected effects.
\vspace*{36pt}

PACS number(s): 98.80.Cq, 14.60.St, 98.80.Ft

\renewcommand{\thefootnote}{\arabic{footnote}}
\addtocounter{footnote}{-2}
\end{quote}
\end{titlepage}

\newpage

\baselineskip=24pt
\renewcommand{\baselinestretch}{1.5}
\footnotesep=14pt

Present experimental bounds to the electromagnetic properties of the
tau neutrino are several orders of magnitude less stringent than the
bounds to the corresponding properties for electron and muon
neutrinos. For instance, while the upper limits to the diagonal
magnetic moments of the electron and muon neutrinos are ($\mu_B$ is a
Bohr magneton) $1.1\times10^{-9}\mu_B$ and $7.4\times10^{-10}\mu_B$
respectively \re{PDB}, the experimental upper limit to the diagonal
magnetic moment of the tau neutrino is $\mu_{\nut} < 5.4 \times
10^{-7}\mu_B$ \re{beamdump}.  More stringent bounds on neutrino
magnetic moments of order $10^{-10}$ to $10^{-12}$ are available from
astrophysical constraints \re{Raffelt}, mainly from the cooling of
stars and from the study of SN 1987A. However, these bounds apply only
if mass of the neutrino species does not exceed the stellar
temperatures relevant for neutrino production. Furthermore,
astrophysical constraints are model dependent since they assume that
the outgoing wrong helicity neutrinos are completely sterile
\re{Babu}.

Big-bang nucleosynthesis (BBN) is a precious tool that has been
employed to constrain many neutrino properties \re{KT}, so it is no
surprise that BBN can be used to bound neutrino magnetic moments.  In
1981, Morgan \re{Morgan} showed that the ``sterile'' right-handed
degree of freedom of Dirac neutrinos\footnote{If the neutrino has a
magnetic moment it must be a Dirac fermion (we do not consider
transitional magnetic moments in this paper).} can be populated
through processes like $e\nu_L \rightarrow e\nu_R$ and $e^-e^+
\rightarrow \nu_R{\bar \nu}_L$, mediated through a virtual photon
coupled to the neutrino through its magnetic moment.  The degree to
which the right-handed neutrino is populated depends upon the strength
of the above interactions, which in turn is proportional to the
magnitude of the neutrino magnetic moment.  Thus, endowing a neutrino
species with a magnetic moment potentially leads to an unacceptable
doubling of the contribution of that species to the energy density,
jeopardizing BBN's successful predictions. If the neutrino species is
relativistic at freeze out, one must require that right-handed
neutrinos decouple before the QCD phase transition so that their
number density is diluted by the huge entropy shift associated with
the transition.  Such a requirement translates into an upper limit to
the neutrino magnetic moment: $\mu_{\nut} \le$ 1 to 2$\times
10^{-11}\mu_B$.  However, in deriving this limit Morgan did not
consider the possibility of non-zero neutrino masses.  There are two
main effects to be considered if one allows a non-zero mass for the
neutrino species in question.  First, the neutrino may not be
relativistic at freeze out, and its number density must be calculated
by solving the Boltzmann equation.  Furthermore, the scaling of the
neutrino energy density with temperature depends upon the neutrino
mass.  Therefore, Morgan's useful limit applies only if the neutrinos
are ultra-relativistic around freeze out and BBN, that is if $m_{\nut}
< 0.1 \MeV$.

The tau neutrino could be heavier than 0.1 MeV;\footnote{We assume
here that the mass of the muon neutrino is less than 0.1 MeV.}  in
fact, the present upper bound to the tau-neutrino mass is $m_{\nut} <
24$ MeV \re{ALEPH}.  Although somewhat model dependent, more stringent
bounds on the neutrino masses can be determined from cosmological
considerations. In particular, if the relic heavy-neutrino energy
density today is sufficiently large, the predicted age of the universe
will be less than observed.  If the neutrino is stable and it is
nonrelativistic today, the age limit ($\Omega h^2<1$) constrains the
mass of any stable neutrino species to be less than the
Cowsik--McClelland limit, $m_\nu < 91.5$ eV \re{CM}. Of course if the
neutrino is unstable, the Cowsik--McClelland limit can be evaded
\re{DKT}. But even in this case there are lifetime-dependent limits to
the neutrino mass.  If the heavy-neutrino lifetime is longer than a
second or so, it can give an additional contribution to the energy
density during nucleosynthesis and spoil the successful predictions of
standard calculations.  Using these kind of considerations, BBN
constraints to the tau-neutrino mass excludes the range $0.3 < m_\nu <
25\;\MeV$ if it is a Dirac fermion, and the range $0.5 < m_\nu <
25\;\MeV$ if it is a Majorana fermion \re{KTCS}.  (It was assumed in
the above BBN analysis that the neutrino eventually decays after BBN;
if it decays after decoupling but before or during BBN, the situation
is more complicated \re{DGK}.)

The Cowsik--McClelland limit and the nucleosynthesis considerations
might be modified if one introduces new interactions that changes the
neutrino annihilation cross section.  This is the case if the neutrino
has a large diagonal magnetic moment, because a large magnetic moment
would increase $\nu$--$\bar{\nu}$ annihilation (creation) into (by)
$e^\pm$, keeping the neutrinos in equilibrium below the canonical
(including only weak processes) neutrino decoupling temperature of
about an MeV.  If the neutrino mass is sufficiently small (much less
than an electron mass) and remains coupled to electrons while the
electrons annihilate, the neutrino number density will be {\it
increased} because part of the electron's entropy will be shared with
the neutrinos.  However, if the neutrino mass is not much less $m_e$
and it remains in equilibrium through magnetic-moment mediated
interactions, its energy density will be Boltzmann suppressed before
decoupling, weakening the BBN constraints.

In this letter we study how the interplay between the neutrino mass
 and magnetic moment modifies the cosmological constraints to
 tau-neutrino properties from the age of the universe and BBN.
 Besides the mere extension of the upper limit on $\mu_{\nut}$ to
 larger neutrino masses, the main purpose of our letter is to give a
 final answer to the intriguing possibility that tau neutrinos with a
 large magnetic moment could form cold dark matter.

Giudice \re{Giudice} first observed that if tau neutrinos are stable,
have a mass in the range $m_{\nut} \sim 1$ to $10\; \MeV$, and are
endowed with a magnetic moment of $\mu_{\nut} \sim 10^{-6}\mu_B$, they
would stay in equilibrium through their magnetic-moment interactions
and would decouple when they are nonrelativistic.  If the magnetic
moment is large enough, their final abundance might give rise to a
universe with $\Omega_\nu h^2 \simeq 1$. Although the latest
experimental upper limit on $\mu_{\nut}$ \re{beamdump} seems
marginally at odds with Giudice's scenario, it is worthwhile to
investigate this hypothesis further \re{BerRub}.

Giudice made use of the fact that Morgan's conclusions about doubling
the effective tau-neutrino number density by populating the
right-handed component can not be applied directly to MeV-mass
neutrinos.  This is because their energy density is Boltzmann
suppressed at freeze-out and during BBN.  Thus, even including the
right-handed components, tau neutrinos will not contribute so much to
the energy density as to spoil BBN.  We show that while this is
approximately true, it is not exactly true.  BBN is such a sensitive
probe of the expansion rate of the universe at the temperatures of
interest that even a small contribution to energy density is
important.  Therefore the contribution of the right-handed neutrino to
BBN requires a careful treatment.  We report the results of such an
investigation in this communication.

There are three effects that must be carefully accounted for: 1) After
a massive neutrino species decouples and becomes nonrelativistic its
energy density grows relative to the energy density of a massless
neutrino species \re{KS}. Although one must solve the Boltzmann
equation to compute the energy density of the heavy neutrino (see
below), it is possible to estimate this effect by observing that
$\rho_\nu(m_\nu \neq 0)/\rho_\nu(m_\nu = 0) \simeq (m_\nu/3.15 T_\nu)r
\propto t^{1/2}$ when $T \sim m_\nu$, where $r$ is the ratio of the
number density of massive neutrinos to massless neutrinos after
freeze-out.  2) A neutrino species with a mass in the MeV range and
with a magnetic moment close to the present experimental limit
decouples when it is semi-relativistic.  Neither the relativistic nor
the nonrelativistic cross section can be used, and a general treatment
of the thermal-averaged annihilation cross section used in the
Boltzmann equation for the neutrino abundance is required. (Giudice
performed his analysis in the extreme nonrelativistic limit.)  3)
Plasma effects must, at least a priori, be considered.  Not only do
thermal corrections to the amplitudes of the main processes involved
in BBN have to be included \re{Dicus}, but more importantly, the mass
corrections due to the electromagnetic coupling of the particles to
the relativistic plasma must be accounted for. For example, the photon
in the thermal bath becomes a {\it plasmon} and acquires an effective
mass \re{Kapusta}.  The plasmon mass has a double effect. In first
place, it affects the electromagnetic channel of the neutrino
annihilation cross section. Although this is a second-order effect
some resonance might enhance it dramatically \re{EKS}. Secondly, a
plasmon mass gives rise to new processes that are kinematically
forbidden in the vacuum. In our case the most relevant of these
processes is the decay $plasmon \rightarrow \nu{\bar \nu}$, having a
rate \re{plasmons}
\be
\Gamma_P = {\mu^2_{\nut}\over
16\pi}\,\left(\omega^2_P - 4 m^2_{\nu}\right)^{3/2}\, {K_1(x_P)\over
K_2(x_P)},
\ee where $x_P = m_P/T$ with $m_P$ the
temperature-dependent plasmon mass, $\omega_P \sim 0.1 T$ is the plasma
frequency, and $K_i(x)$ are the modified
Bessel functions of order $i$.  However,  the threshold
$\omega_{P} \geq 2m_{\nu}$ reduces the importance of the process
$plasmon \rightarrow \nu{\bar \nu}$ during BBN for MeV-mass  neutrinos.
Analogously, since $2m_{\nu} > \omega_P$, screening effects
induced by the plasma on the photon propagator turn out to be
negligible.  The relative unimportance of these considerations were verified
by directly including them in our numerical calculations.

The  cross section for the electromagnetic channel of the
process $\nu{\bar \nu} \rightarrow  e^-e^+$ is
\be
\sigma_{\nu{\bar \nu} \rightarrow e^-e^+} =  {\alpha \mu_\nu^2 \over 6}
\left({\displaystyle {1 - 4 m_e^2/ s}\over {1 - {4 m_\nu^2 /  s}}}
\right)^{1/2}
 \left(1 + 8\,{m^2_\nu \over s} + 2\,{m^2_e \over s} +
16\,{m^2_\nu m^2_e\over s^2}\right) , \label{sigma}
\ee
where $\sqrt{s}>2m_e$ is the total center-of-mass energy.

The weak contribution to the annihilation process of Eq.\
(\ref{sigma}) can be neglected if the neutrino magnetic moment is
larger than $10^{-10} (m_{\nu}/1\,\MeV)\,\mu_B$.  We will work within
the limits of this assumption.\footnote{If the magnetic moment is
larger than $10^{-10} (m_\nu/1~{\rm MeV})\mu_B$, then neutrino
annihilation will occur predominantly through photon exchange, rather
than $Z$ exchange.  For $m_\nu \la 100$ keV, considerations of stellar
energy loss by neutrino pair emission limits the magnetic moment to be
greater than about $10^{-11}\mu_B$.  Thus, we will consider neutrinos
more massive than 100 keV.}

Because both helicity eigenstates of the neutrino are symmetric with
respect to electromagnetic interactions, we do not differentiate
between them in our calculations.  For this reason the processes
$e{\nu}_{L(R)} \longleftrightarrow e{\nu}_{R(L)}$ changes neither the
total, nor the relative ${\nu}_L$ vs.\ ${\nu}_R$
abundances.\footnote{The $\nu_{L(R)}$ helicity eigenstates should not
be confused with the chirality eigenstates. Since we ignore weak
interactions, chirality does not play a role in our analysis.}

The Boltzmann equation for the abundance of the heavy neutrino is
\re{Gelmini}
\be {dY\over dx} =
-\left(\frac{\pi}{45}\right)^{1/2}\;{g_*^{1/2} m_{\nu} m_{Pl} \over
x^2 }\, \langle \sigma v_{\mbox{\small{M{\o}l}}} \rangle \,\left(Y^2 -
Y^2_{EQ}\right),
\label{Boltz}
\ee
where $x = m_{\nu}/T$, $Y = n_{\nu}/s$ is the ratio of the $\nut$
number density to the total entropy density of the universe,
$v_{\mbox{\small{M{\o}l}}}$ is the M{\o}ller invariant flux factor,
and $m_{Pl}=G_N^{-1/2}$ is the Planck mass. The parameter $g_*$ is
defined as
\be g^{1/2}_* = {h_{eff}\over g^{1/2}_{eff}}\,\left(1 +
{1\over 3}\, {T\over h_{eff}(T)}\, {dh_{eff}(T)\over dT} \right) ,
\ee
where the effective number of degrees of freedom for the energy
density, $g_{eff}(T)$, and for the entropy density, $h_{eff}(T)$, are
defined as
\be \rho = g_{eff}(T)\,{\pi^2\over 30}\,T^4; \qquad s=
h_{eff}(T)\,{2\pi^2\over 45}\,T^3\ .
\ee
Following \re{Gelmini}, the
thermal averaged cross section times the M{\o}ller velocity is
\be
\langle \sigma v_{\mbox{\small{M{\o}l}}}\rangle  =
{1\over 8 m^4_{\nu} T K^2_2(x)}
\int^\infty_{4m^2_{\nu}} \sigma(s)\,
(s - 4m^2_{\nu})\sqrt{s} K_1(\sqrt{s}/T) \,
ds.
\label{sigmav}
\ee
We have used the Maxwell--Boltzmann distribution to compute the
thermal-averaged cross section (for a detailed review of computations
in this approximation see e.g., Ref.\ \re{Wolfram}).  Although
normally this is a very good approximation only for temperatures $T
\undersim{<} 3 m_\nu$, we have checked that at the freeze-out
temperature (the only temperature around which Eq.\ (\ref{sigmav})
plays a relevant role) the approximation is adequate.

The neutrino decoupling temperature $T_F$ is here defined by the condition
$Y(T_F) - Y_{EQ}(T_F) = 1.5 Y_{EQ}$, where
\be
Y_{EQ} = {n_\nu^{EQ}\over s} = {45\over \pi^4}\, {I_\nu(x)\over h_{eff}(T)},
\ee
with
\be
I_\nu(x) = \int_1^{\infty} dz z {\sqrt{z^2 - x^2}\over e^z + 1} \ .
\ee

\begin{figure}[t]
\centerline{\hbox{\psfig{figure=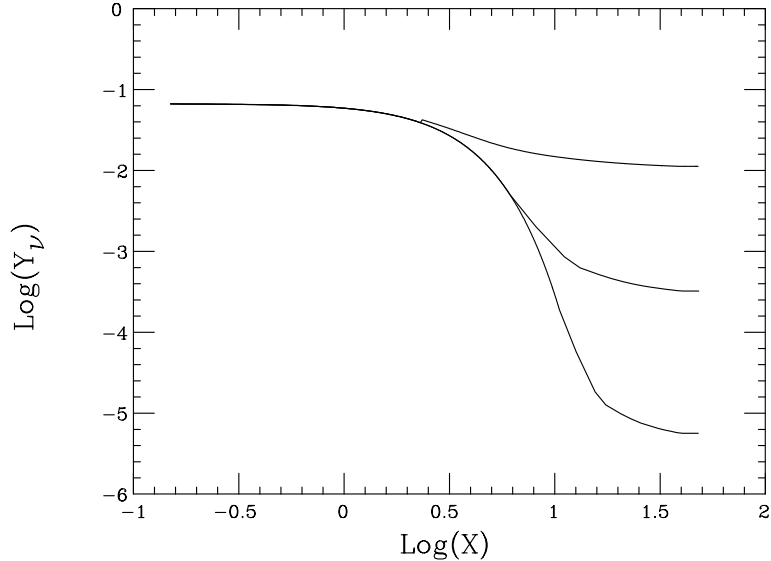,width=12cm}}}
\label{fig:Fig. 1}
 \caption{ Tau neutrino abundance vs.\
	   the parameter $x=m_{\nut}/T$ is represented for different value of
           $\mu_{\nut}$: from below, the three different curves refer to
           $\mu_{\nu} = 10^{-6}\mu_B,\ 10^{-7}\mu_B,$ and $10^{-8}\mu_B$.
           Here we have chosen $m_{\nu} = 1\;\MeV$.  The logarithms are base
10.}
\end{figure}

We numerically solved Eq.\ (\ref{Boltz}) to compute the tau-neutrino
abundance as function of $T$ for fixed values of $m_{\nut}$ and
$\mu_{\nut}$. Since the freeze-out temperature increases with
$\mu_{\nut}$, it is natural to expect that the final tau-neutrino
abundance is suppressed as the magnetic moment increases. This is
clearly visible in Fig.\ 1.  Assuming the tau neutrinos to be stable
we can easily check their effect on the dynamics of the universe for
several values of $m_{\nut}$ and $\mu_{\nut}$.  In particular, we
first consider the contribution to the present energy density of the
universe due to massive tau neutrinos:
\begin{equation}
\Omega_{\nut}h^2 \equiv {\rho_{\nut \, 0}\over \rho_C} h^2 =
{m_{\nut} s_0 Y_{\nut \, 0}\over 1.054 \;\MeV{\rm cm}^{-3}},
 \label{omega}
\end{equation}
where $0$ indicates quantities evaluated at the present time.
Requiring $\Omega_{\nut} h^2 \le 1$, we can verify which region of the
parameter space $\mu_{\nut}$ versus $m_{\nut}$ is compatible with the
age constraint.

Of course if the tau neutrino is unstable, the cosmological age
constraint discussed above does not apply. However we can still use
BBN to limit the properties of the tau neutrino provided that the
lifetime, $\tau_{\nut}$, is greater than about a second.

\begin{figure}[t]
\centerline{\hbox{\psfig{figure=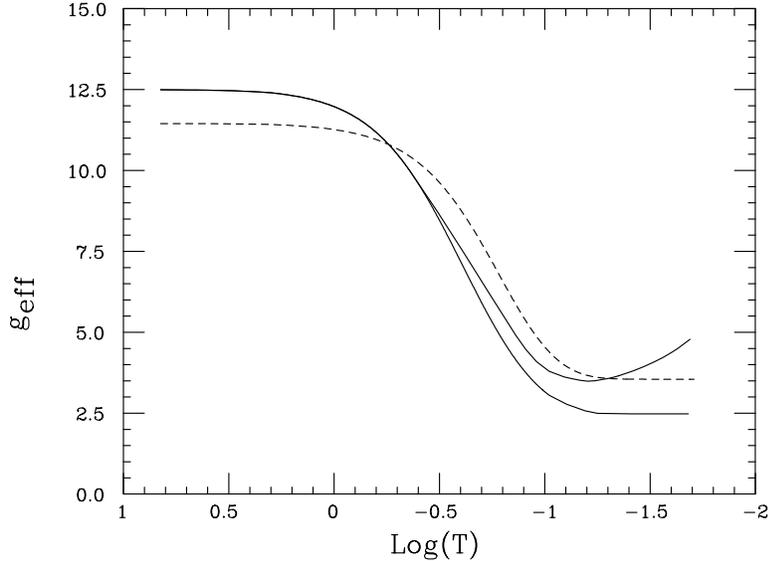,width=12cm}}}
\label{fig:Fig. 2}
 \caption{The effective number of degrees of freedom in the energy
	density is shown as function of temperature for two chosen
	value of $\mu_{\nu}$ and $m_{\nu} = 1$ MeV.  The upper solid
        curve is for
	$\mu_{\nut} = 10^{-8}\mu_B$; the lower solid curve is for
	$\mu_{\nut} = 10^{-6}\mu_B$. The reader can compare our result
	with the result obtained for 3.4 standard massless neutrinos
	shown by the dashed line. The logarithm is base 10.}
\end{figure}

To evaluate the impact of the massive tau neutrino with a large
magnetic moment on BBN we must know how it modifies the effective
number of degrees of freedom of the energy density, $g_{eff}(T)$, for
$0.1 \undersim{<} T \undersim{<} 10$ MeV, since the light element
relic abundances depend critically on the expansion rate of the
universe during BBN, which in turn is parameterized by $g_{eff}$
\re{KT}:
\begin{equation}
H(T) = 1.66 g_{eff}^{1/2}(T) {T^2\over m_{Pl}} \ .
\end{equation}
The tau neutrino contribution to $g_{eff}$ is given by
\begin{equation}
g_{\nut}(T) = \rho_{\nut}(T) \left({30\over \pi^2}\right) T^{-4},
\end{equation}
where $\rho_{\nut} = s Y_{\nut} \sqrt{(3.15 T_{\nut})^2 +
m_{\nut}^2}$\,, and $T_{\nut}$ is computed by imposing entropy
conservation.  Fig.\ 2 clearly demonstrates that $g_{eff}$ grows as
the decoupled tau neutrinos become nonrelativistic.  This effect
becomes less pronounced as the magnetic moment is increased.  Of
course this is simply because increasing $\mu_{\nut}$ decreases $T_F$,
leading to a tau-neutrino energy density more effectively Boltzmann
suppressed before freeze out.  For this reason, values of the
tau-neutrino magnetic moment larger than $10^{-8} \mu_B$ are not
expected to have a large effect on BBN if $m_{\nut} > 0.1 \ \MeV$.

\begin{figure}[t]
\centerline{\hbox{\psfig{figure=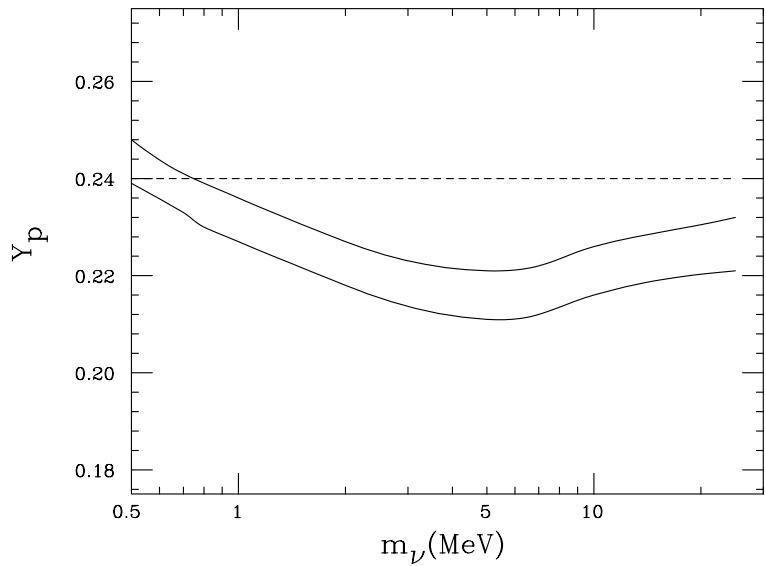,width=12cm}}}
\label{fig:Fig. 3}
 \caption{The predicted $^4{\rm He}$ relic abundance is represented
          as function of the tau-neutrino mass for two values of
          $\mu_{\nu}$. The upper curve is for $\mu_{\nut} = 10^{-8}\mu_B$,
while
          the lower one is for  $\mu_{\nut} = 10^{-6}\mu_B$. The dashed line
          corresponds to the observational upper limit.}
\end{figure}

To check this in detail and in order to be able to evaluate the
effects of the neutrino mass and magnetic moment on light element
production, we incorporated our results for the abundance as a
function of temperature into the standard nucleosynthesis code
\re{Kawano}.  In Fig.\ 3 our predictions for the relic $^4{\rm He}$
abundance as a function of the tau-neutrino mass are shown for two
values of the magnetic moment. As expected, the predicted abundance
$Y_P$ is suppressed with increasing $\mu_{\nut}$. Increasing the mass
above a few MeV increases $Y_P$, since the tau neutrinos then become
nonrelativistic earlier.  For small masses, $Y_P$ grows with
decreasing $m_{\nut}$. This is due both to the less effective
Boltzmann suppression and to the entropy transfer from $e^\pm$
annihilation to the tau neutrinos.

In order to discriminate which region of the $m_{\nu}$ versus\
$\mu_{\nu}$ parameter space is compatible with observations, we
require that the predicted light element abundances do not exceed the
observational limits \re{Olive}: $Y_P \le 0.24$; $({\rm D} + {\rm
^3He})/{\rm H} \le 1.1\,\times 10^{-4}$; and $^7{\rm Li}/{\rm H} \le
1.7\times\, 10^{-10}$.  Since the baryon--to--photon ratio $\eta$ is a
free parameter, for every chosen pair of $m_{\nut}$ and $\mu_{\nut}$
we fix it at the minimum value compatible with the $({\rm D} + {\rm
^3He})/{\rm H}$ upper limit. Then we check if the predicted $^4{\rm
He}$ relic abundance is consistent with the upper limit of 0.24. The
$^7{\rm Li}$ constraints turn out to be always less stringent than the
limits coming from $^4{\rm He}$.

\begin{figure}[t]
\centerline{\hbox{\psfig{figure=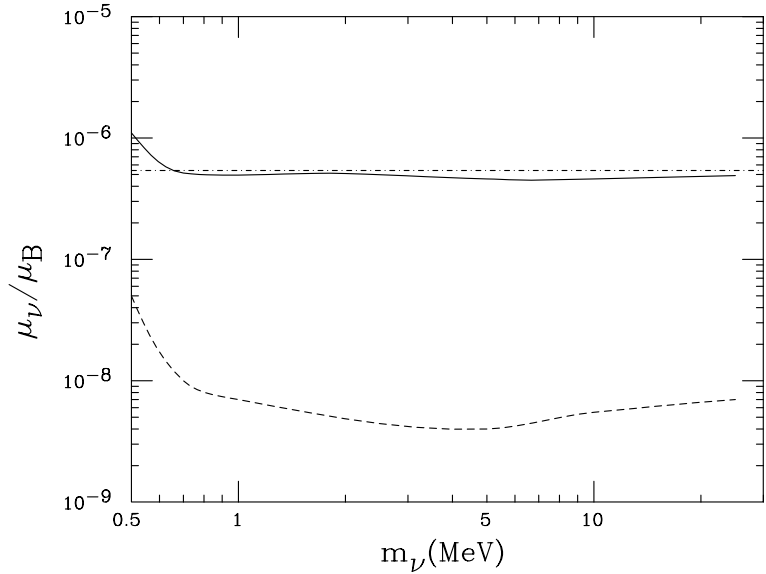,width=12cm}}}
\label{fig:Fig. 4}
 \caption{Exclusion plot in the tau neutrino magnetic moment---mass
          parameter space. The solid  line provides the lower limit
          to $\mu_{\nut}$ coming from the requirement $\Omega h^2 \le 1$.
          The dashed line provides the corresponding limit from BBN
          considerations. The dotted-dashed line represent the experimental
          upper limit. }
\end{figure}

Our results are summarized in Fig. 4. As the reader can observe, the
age-based constraints are much more stringent than BBN constraints if
the tau neutrino is stable. In this case the border between the
allowed and forbidden regions in the $m_{\nut}$ versus.\ $\mu_{\nut}$
parameter space from the age constraint almost coincides with the
experimental limit line. This is a remarkable coincidence. In fact,
the age limits are {\it lower} limits to $\mu_{\nut}$, whereas the
experimental limits are {\it upper} limits.  This means that nearly
the entire parameter space for $0.1 \MeV \undersim{<} m_{\nut}$ and
$10^{-10}(m_{\nut}/1 \MeV) \undersim{<} \mu_{\nut}$ (the very range to
which our consideration apply) is excluded by our considerations.  To
be precise, a very small region between the experimental and the
age-based constraints remains open.  Even this region would be closed
using a slightly larger value of $h$ as recent observations suggest.

As a consequence, Giudice's hypothesis is definitely ruled out.
Furthermore, our results improve the upper limit on the tau-neutrino
magnetic moment by several orders of magnitude in the mass range we
considered.  We have checked that plasma-physics effects are
subdominant.  We have to stress that this limit is valid only if the
tau neutrino is stable, as indeed Giudice assumed. Stability can be
achieved by imposing some additional symmetries, e.g., individual
lepton-number conservation.  Of course, in any case some new physics
beyond the standard model must be introduced in order to have such a
large value of $\mu_{\nut}$. Furthermore, a tau neutrino with mass
larger than 1.1 MeV decaying according to the a minimally extended
standard model via the channel $\nut \rightarrow \nu_e e^+e^-$ is
incompatible with BBN.  In fact, since experimental data constrains
the $\nut$ lifetime to be 1 s$\le \tau_{\nut} \le 10$ s, if $m_{\nut}
> 1.1$ MeV, electrons and positrons produced from this decay would
induce the photodestruction of light elements \re{photod}.

If the tau neutrino is unstable but the lifetime exceeds one second, a
band of magnetic moment values, roughly $10^{-8}\mu_B \undersim{<}
\mu_{\nut} \undersim{<} 10^{-6} \mu_B$, remains compatible with
experimental and cosmological bounds. This confirms the result of
Ref.\ \re{KFMS} and extends it to a wider tau-neutrino mass range. It
is understood that in this case the tau neutrino has to decay in some
non-standard way in order the decay products do not affect
dramatically the light element relic abundances.

\vspace{36pt}

\centerline{\bf ACKNOWLEDGMENTS}

The work of D.\ G.\ was supported in part by Istituto
Nazionale di Fisica Nucleare, Sezione di Roma, and by the EEC
contract SC1*-CT91-0650.  The work of E.\ W.\
K.\ was supported in part by the Department of Energy and NASA (grant
NAG5--2788).

\frenchspacing
\def\prpts#1#2#3{Phys. Reports {\bf #1}, #2 (#3)}
\def\prl#1#2#3{Phys. Rev. Lett. {\bf #1}, #2 (#3)}
\def\prd#1#2#3{Phys. Rev. D {\bf #1}, #2 (#3)}
\def\plb#1#2#3{Phys. Lett. {\bf #1B}, #2 (#3)}
\def\npb#1#2#3{Nucl. Phys. {\bf B#1}, #2 (#3)}
\def\apj#1#2#3{Astrophys. J. {\bf #1}, #2 (#3)}
\def\apjl#1#2#3{Astrophys. J. Lett. {\bf #1}, #2 (#3)}
\begin{picture}(400,50)(0,0)
\put (50,0){\line(350,0){300}}
\end{picture}

\vspace{0.25in}

\def\labelenumi{[\theenumi]}

\begin{enumerate}

\item\label{PDB} Particle Data Group, \prd{50}{1173}{1995}.

\item\label{beamdump} A. M. Cooper-Sarkar et al., \plb{280}{153}{1992}.

\item\label{Raffelt} G. Raffelt, \prl{64}{2856}{1990};
G. Raffelt, D. Dearborn, and J. Silk, \apj {336}{61}{1989}.

\item\label{Babu} K. S. Babu, R. N. Mohapatra, and I. Z. Rothstein,
\prd{45}{3312}{1992}.

\item\label{KT} E. W. Kolb and M. S. Turner, {\it The Early Universe},
(Addison-Wesley, Redwood City, 1990).

\item\label{Morgan} J.A. Morgan, \plb{102}{247}{1981}.

\item\label{ALEPH} ALEPH collab., CERN PPE/95-03.

\item\label{CM} S. S. Gerstein and Ya. B. Zeldovich, JEPT Lett.
{\bf 4}, 174 (1966); R. Cowsik and J. McClelland, \prl{29}{669}{1972}.

\item\label{DKT} D. A. Dicus, E. W. Kolb, and V. L. Teplitz,
\prl{39}{168}{1977}.

\item\label{KTCS} E. W. Kolb, M. S. Turner, A. Chakravorty, and D.  N. Schramm,
\prl {67}{533}{1991}; A. Dolgov and I. Rothstein, \prl{71}{476}{1993}.

\item\label{DGK} S. Dodelson, G. Gyuk, and M. S. Turner,
\prl {72}{3754}{1994}.

\item\label{Giudice} G. Giudice, \plb{251}{460}{1990}.

\item\label{BerRub} Concerning some other experimental consequences
of a tau neutrino having such properties, see also:
L. Bergstr\"om and H. R. Rubinstein, \plb{253}{168}{1991};
D. Grasso and M. Lusignoli, \plb{279}{161}{1992}.

\item\label{KS} E. W. Kolb and R. J. Scherrer, \prd{25}{1481}{1982}.

\item\label{Dicus} D. A. Dicus et al., \prd{26}{2694}{1982}.

\item\label{Kapusta} J. I. Kapusta, {\it Finite Temperature Field Theory},
(Cambridge University Press, 1989).

\item\label{EKS} K. Enqvist, K. Kainulainen, and V. Semikoz,
\npb{374}{392}{1992}.

\item\label{plasmons} E. Braaten and D. Segel, \prd{48}{1478}{1993};
D. Grasso and E. W. Kolb, \prd{48}{3522}{1993}.

\item\label{Gelmini} P. Gondolo and G. Gelmini, \npb{360}{145}{1991}.

\item\label{Wolfram} E. W. Kolb and S. Wolfram, \npb{172}{224}{1980}.

\item\label{Kawano} L. Kawano, {\it Let's Go Early Universe: Guide to
Primordial Nucleosynthesis Programming}, FERMILAB-PUB-88/34-A.
This code is a modernized and optimized version of the code written by
R. V. Wagoner, \apj {179}{343}{1973}.

\item\label{Olive} K. Olive et al., \plb{236}{454}{1990}.

\item\label{photod} N. Teresawa, M. Kawasaki, and K. Sato,
\npb{302}{697}{1988}.

\item\label{KFMS} L. H. Kawano, G. M. Fuller, R. A. Malaney, and M. J. Savage,
\plb{275}{487}{1992}.

\end{enumerate}

\end{document}